# Novel Super-Resolution Method Based on High Order Nonlocal-Means

Kang Yong-Rim, Kim Yong-Jin

*Department of Physics, Kim Il Sung University, Pyongyang, DPRK*

## Abstract

Super-resolution without explicit sub-pixel motion estimation is a very active subject of image reconstruction containing general motion. The Non-Local Means (NLM) method is a simple image reconstruction method without explicit motion estimation. In this paper we generalize NLM method to higher orders using kernel regression can apply to super-resolution reconstruction. The performance of the generalized method is compared with other methods.

## 1. Introduction

Super-resolution is a technology that combines several low-resolution images into one or more high-resolution results. This problem consists of denoising, deblurring, and scaling-up tasks [1, 2].

In the classic super-resolution algorithms, the motion in the sequence is estimated by sub-pixel accuracy, and then input images are combined according to the motion vectors [1, 2, 3]. The most important step is the accurate motion estimation. Super-resolution is possible only when sub-pixel motion estimation is available [1]. When the motion estimation is inaccurate, annoying ringing artifacts appear in the results, making it even worse than low-resolution inputs. This motion estimation problem is discussed in [4, 5, 6].

In the classic super-resolution algorithms, only global motion is treated. In the case of local motion, accurate motion estimation is very difficult. It means that general movies that contain local motion cannot be handled well by classical super-resolution methods.

Several super-resolution methods without explicit motion estimation were developed recently. Nonlocal-means method [7] is an image denoising method does not require motion estimation. Protter et al.[8] generalized this method to super-resolution problem. A super resolution method without explicit sub-pixel motion estimation using 3D kernel regression was proposed by Takeda et al.[9].

In this paper, we show that NLM method can be thought as zeroth order kernel regression and generalize it to higher orders.

## 2. Non-Local Means Filter in Denoising

The NLM filter is based on the assumption that image content can be repeated within some neighborhood [7]. Each pixel's denoised value is calculated by weighted average of all pixels in its neighborhood. The weight of each pixel in the relevant neighborhood reflects the probability that this pixel and the center pixel had the same value in the original image. These filters are described by

formula:

$$\hat{\mathbf{x}}[k,l] = \frac{\sum_{(i,j)\in N(k,l)} w[k,l,i,j]\mathbf{y}[i,j]}{\sum_{(i,j)\in N(k,l)} w[k,l,i,j]} \quad (1)$$

where $N(k,l)$ stands for the neighborhood of the pixel $(k,l)$ and the term $w[k,l,i,j]$ is the weight for the $(i,j)$-th neighbor pixel. The input pixels are $\mathbf{y}[k,l]$ and the output result in that position is $\hat{\mathbf{x}}[k,l]$.

NLM filter is closely related to the bilateral filter, where weights are computed based both on radiometric similarity and geometric similarity between the pixels. The radiometric part in the weights of the NLM is computed by computing the Euclidean distance between two image patches centered at these two pixels. Defining $\hat{\mathbf{R}}_{k,l}$ as an operator that extracts a patch of a fixed and predetermined size $(q \times q)$ from an image, the NLM weights are given by

$$w[k,l,i,j] = \exp\left\{-\frac{\left\|\hat{\mathbf{R}}_{k,l}\mathbf{y} - \hat{\mathbf{R}}_{i,j}\mathbf{y}\right\|_2^2}{2\sigma_r^2}\right\} \cdot f\left(\sqrt{(k-i)^2 + (l-j)^2}\right) \quad (2)$$

The function $f$ reflects the geometric distance, and it is monotonically non-increasing [7].

The NLM filter is generalized to the super-resolution problem in [8]. The first step is a separation of the deblurring from the fusion of the images. Using the substitution $\mathbf{Z} = \mathbf{HX}$, the first problem is the estimation of $\mathbf{Z}$ from the measurements $\{\mathbf{y}_t\}_{t=1}^T$, by minimizing the energy function

$$\eta_{SR}^A(\mathbf{Z}) = \sum_{(k,l)\in\Omega} \sum_{t\in[1,...,T]} \sum_{(i,j)\in N^L(k,l)} w[i,j,k,l,t] \times \left\|\mathbf{D}_p \mathbf{R}_{k,l}^H \mathbf{Z} - \mathbf{R}_{i,j}^L \mathbf{y}_t\right\|_2^2. \quad (3)$$

Then we estimate $\mathbf{X}$ using $\hat{\mathbf{Z}}$ to by minimizing

$$\eta_{SR}^B(\mathbf{X}) = \left\|\hat{\mathbf{Z}} - \mathbf{HX}\right\|_2^2 + \lambda TV(\mathbf{X}). \quad (4)$$

From Eq. (2) it can be seen that the method tends to find a single constant at each pixel which minimizes the reconstruction error of the patch under consideration. The weights detect similar patches and ensure that the cost function penalizes only the error with respect to pixels with similar neighborhoods. In general, the NLM method assumes that an image can be modeled to be locally constant. This limits the performance of NLM. Therefore, higher-order approximation is required to handle regions of finer detail.

## 3. Higher Order Non-Local Means via Kernel Regression

In this section we generalize the Non-Local Means method to higher order. We start from the

kernel regression method [10]. Measured data is given by

$$y_i = z(\mathbf{x}_i) + \varepsilon_i, \qquad i = 1, 2, \ldots, P \qquad (5)$$

where $z(\cdot)$ is the unspecified regression function and $\varepsilon_i$ s are the independent and identically distributed zero mean noise values. If we assume that $z(\mathbf{x}_i)$ is locally smooth to some order $N$ we can use local expansion to estimate a value at a point $\mathbf{x}$. If $\mathbf{x}$ is near the sample at $\mathbf{x}_i$, we have the N-term Taylor series

$$\begin{aligned}z(\mathbf{x}_i) &= z(\mathbf{x}) + \{\nabla z(\mathbf{x})\}^T (\mathbf{x}_i - \mathbf{x}) + \frac{1}{2}(\mathbf{x}_i - \mathbf{x})^T \{Hz(\mathbf{x})\}(\mathbf{x}_i - \mathbf{x}) + \cdots \\ &= \beta_0 + \beta_1^T (\mathbf{x}_i - \mathbf{x}) + \beta_2^T \text{vech}\{(\mathbf{x}_i - \mathbf{x})(\mathbf{x}_i - \mathbf{x})^T\} + \cdots\end{aligned} \qquad (6)$$

where $\nabla$ and $H$ are gradient $(2 \times 1)$ and Hessian $(2 \times 2)$ operators and vech($\cdot$) is half-vectorization operator of the "lower-triangular" portion of a symmetric matrix. The $\beta_n$ s are computed from the following optimization problem:

$$\min_{\{\beta_n\}} \sum_{i=1}^{P} \left[ y_i - \beta_0 - \beta_1^T (\mathbf{x}_i - \mathbf{x}) - \beta_2^T \text{vech}\{(\mathbf{x}_i - \mathbf{x})(\mathbf{x}_i - \mathbf{x})^T\} - \cdots \right]^2 K_\mathbf{H}(\mathbf{x}_i - \mathbf{x}) \qquad (7)$$

with

$$K_\mathbf{H}(\mathbf{t}) = \frac{1}{\det(\mathbf{H})} K(\mathbf{H}^{-1}\mathbf{t}) \qquad (8)$$

where $K$ is the 2-D kernel function, and $\mathbf{H}$ is the $2 \times 2$ smoothing matrix. It is possible to express it in a matrix form as a weighted least-squares optimization problem

$$\begin{aligned}\hat{\mathbf{b}} &= \arg\min_{\mathbf{b}} \|\mathbf{y} - \mathbf{X}_\mathbf{X}\mathbf{b}\|_{\mathbf{W}_\mathbf{X}}^2 \\ &= \arg\min_{\mathbf{b}} (\mathbf{y} - \mathbf{X}_\mathbf{X}\mathbf{b})^T \mathbf{W}_\mathbf{X} (\mathbf{y} - \mathbf{X}_\mathbf{X}\mathbf{b})\end{aligned} \qquad (9)$$

where

$$\mathbf{y} = [y_1, y_2, \cdots, y_P]^T, \qquad \mathbf{b} = [\beta_0, \beta_1^T, \cdots, \beta_N^T]^T \qquad (10)$$

$$\mathbf{W}_\mathbf{X} = \text{diag}[K_\mathbf{H}(\mathbf{x}_1 - \mathbf{x}), K_\mathbf{H}(\mathbf{x}_2 - \mathbf{x}), \cdots, K_\mathbf{H}(\mathbf{x}_P - \mathbf{x})] \qquad (11)$$

$$\mathbf{X}_\mathbf{X} = \begin{bmatrix} 1 & (\mathbf{x}_1 - \mathbf{x})^T & \text{vech}\{(\mathbf{x}_1 - \mathbf{x})(\mathbf{x}_1 - \mathbf{x})^T\} & \cdots \\ 1 & (\mathbf{x}_2 - \mathbf{x})^T & \text{vech}\{(\mathbf{x}_2 - \mathbf{x})(\mathbf{x}_2 - \mathbf{x})^T\} & \cdots \\ \vdots & \vdots & \vdots & \vdots \\ 1 & (\mathbf{x}_P - \mathbf{x})^T & \text{vech}\{(\mathbf{x}_P - \mathbf{x})(\mathbf{x}_P - \mathbf{x})^T\} & \cdots \end{bmatrix} \qquad (12)$$

To compute an estimate of the pixel values, only parameter $\beta_0$ is needed. Therefore, the

least-squares estimation is simplified to

$$\hat{z}(\mathbf{x}) = \hat{\beta}_0 = \mathbf{e}_1^T (\mathbf{X}_\mathbf{X}^T \mathbf{W}_\mathbf{X} \mathbf{X}_\mathbf{X}) \mathbf{X}_\mathbf{X}^T \mathbf{W}_\mathbf{X} \mathbf{y} \tag{13}$$

It can be seen that the NLM method is a special case of the kernel regression. In the zeroth order of kernel regression, it becomes to the NLM formulation. When data points $y_i, i = 1, 2, \cdots P$ are given, normal NLM result in weighted least-square formulation becomes as follows:

$$\hat{\beta}_0 = \hat{z}(\mathbf{x}_i) = \frac{\sum_{i=1}^{P} y_i K_{NLM}(\mathbf{x}_i - \mathbf{x}, y_i - y)}{\sum_{i=1}^{P} K_{NLM}(\mathbf{x}_i - \mathbf{x}, y_i - y)} \tag{14}$$

The method of NLM can be generalized to any arbitrary order combining Eq.(13) and Eq.(14). Combining the advantage of block matching in NLM and higher order property of kernel regression. In this paper, we work with higher orders of 1 and 2.

## 4. Results and Conclusion

We tested the proposed high-order algorithm on real sequences, containing general motion patterns. We chose "Miss America," "Foreman", and "Suzie" just as in [8]. The similarity block size, search area and kernel size are chosen the same as in [8].

Mean-PSNR values of the three sequences are shown in Table I. Results of NLM method [8] and high-order(1st order and 2nd order) methods as well as Lanczos interpolation method are shown.

Table 1. Mean-PSNR results for the three test sequences

| Sequence | Lanszos | NLM (1st it.)[8] | NLM (2nd it.)[8] | Our result (1st order) | Our result (2nd order) |
|---|---|---|---|---|---|
| Miss-America | 34.08 | 35.91 | 34.97 | 36.05 | 37.17 |
| Foreman | 31.25 | 34.27 | 33.13 | 34.45 | 35.84 |
| Suzie | 31.40 | 33.74 | 33.32 | 34.63 | 36.18 |

High-order method can overcome the limit of local consistency assumption of NLM. In all cases, regression of order 2 in fact results in superior performance than first order regression. We find that generalized high-order NLM method shows better performance both visually and in terms of PSNR, and well suited to the super-resolution with general motion pattern.